\documentclass[aps,prb,showpacs,preprintnumbers,psfrac,twocolumn]{revtex4}
\usepackage{amsmath}
\usepackage{epsfig}
\usepackage{graphicx}
\usepackage{dcolumn}
\usepackage{bm}

\renewcommand{\a}{\alpha}

\newcommand{\s}{\sigma}

\def\gl{\lower.35em\hbox{$\stackrel{\textstyle>}{\textstyle<}$}}

\def\gapp{\lower.35em\hbox{$\stackrel{\textstyle>}{\sim}$}}
\def\lapp{\lower.35em\hbox{$\stackrel{\textstyle<}{\sim}$}}

\newcommand{\beq}{\begin{equation}} 
\newcommand{\eeq}{\end{equation}} 
\newcommand{\beqa}{\begin{eqnarray}} 
\newcommand{\eeqa}{\end{eqnarray}}

\newcommand{\om}{\omega}

\def\simleq{\; \raise0.3ex\hbox{$<$\kern-0.75em
      \raise-1.1ex\hbox{$\sim$}}\; }
\def\simgeq{\; \raise0.3ex\hbox{$>$\kern-0.75em
      \raise-1.1ex\hbox{$\sim$}}\; }

  \def\y{\'\i}

\def\non{\nonumber }
\def\beq{\begin{equation} }
\def\eeq{\end{equation} }
\def\beqa{\begin{eqnarray}}
\def\eeqa{\end{eqnarray}}

\def\a{\alpha }

\def\s{\sigma}
\def\med{\frac{1}{2}}
\def\qua{\frac{1}{4}}

\begin{document}
\title{Existence of bulk chiral fermions and crystal symmetry}
\author{J. L. Ma\~nes}
\affiliation{Departamento de F\'{\i}sica  de la Materia Condensada\\
Universidad del Pa\'{\i}s Vasco,
Apdo.~644, E-48080 Bilbao, Spain}
\date{\today}

\begin{abstract}

We consider the existence of bulk chiral fermions 
around points of symmetry 
 in the Brillouin zone of  nonmagnetic $3D$ crystals   
with negligible spin-orbit interactions. We use  group theory   to show  that this is possible, but only for a reduced  number of space groups and   points  of  symmetry   that we tabulate. 
Moreover, we show that for a handful of space groups  
 the existence of bulk chiral fermions  is  not only possible but unavoidable, irrespective of the concrete crystal structure.  
Thus our tables can be used to look for bulk chiral fermions in a specific class of systems, namely that of nonmagnetic $3D$ crystals with  sufficiently weak spin-orbit coupling.  We also  discuss the effects of  spin-orbit interactions 
and possible extensions of our approach to
 Weyl semimetals,  crystals with magnetic order, and systems with Dirac points with pseudospin $1$ and $3/2$. A simple tight-binding  model is used to illustrate some of the issues.
\end{abstract}

 \maketitle


%







\section{Introduction}

Electrons  in the vicinity of the $K$ points in graphene~\cite{novo1,novo2} have linear dispersion relations and behave like massless chiral  fermions. More concretely, the dynamics of electrons  around these points is governed by the relativistic, two-dimensional    massless Dirac  hamiltonian \hbox{$H_0\sim \s_x k_x +\s_y k_y$}, and  many of the exotic electronic properties of graphene stem from this fact.~\cite{electr} This also turns graphene into a potential laboratory for two-dimensional relativistic dynamics, incorporating massless fermions, gauge fields and  curved gravitational backgrounds.~\cite{gauge} Moreover, optical lattices that could be used to simulate  relativistic systems with trapped cold atoms can be fashioned after graphene,~\cite{cold} with control over the properties of the system.~\cite{eng1,eng2}. 
Obviously  three-dimensional analogs of graphene are potentially very interesting. 

Strictly speaking, the massless Dirac hamiltonian $H_0$ describes only the low energy, {\it orbital} dynamics of electrons in graphene. As reviewed in Section~III, spin-orbit coupling in graphene gives fermions a very small mass.~\cite{km} This mass is so small that for most practical purposes spin and orbital degrees of freedom decouple and $H_0$ provides an effective description of an enormous variety of phenomena.~\cite{electr}  

In this paper we will consider $3D$ analogs of graphene, i.e.,  crystals with orbital electron dynamics governed by the $3D$ two-component massles Dirac hamiltonian  \hbox{$H_0\sim v\vec \s\cdot\vec k$}, also known as the Weyl hamiltonian. This hamiltonian describes  massless  chiral fermions, right-handed for $v>0$, left-handed otherwise. Henceforth we will use  the term `orbital Weyl  point' to refer to   points around which the low energy dynamics in the absence of spin-orbit couplings is   described by the $3D$ Weyl hamiltonian, with an additional twofold degeneray due to electron spin. We will also speak of `bulk chiral fermions', keeping in mind that, as in graphene,~\cite{electr} they are exactly  chiral in the limit of vanishing spin-orbit interactions. Also note that it is {\it pseudospin}, not electron spin, which is parallel (or antiparallel) to  $\vec k$ in the chiral limit and that, in these systems, pseudospin is purely orbital in 
origin.

These should be distinguished  from  other systems  where  the {\it total} hamiltonian, including electron spin and spin-orbit interactions,  adopts the form of the  Weyl hamiltonian. These include surface states in topological insulators~\cite{rmp} as well as novel three-dimensional `Weyl semimetals'.~\cite{mc,flux,arc,nhall,tune,trans} The spectrum of these systems,
unlike graphene and its $3D$ analogs considered in this paper,  
 remains gapless  for arbitrary  values of the spin-orbit couplings. Actually, some Weyl semimetals have  \textit{strong}  spin-orbit interactions.~\cite{arc} Possible extensions of our methods to Weyl semimetals will be considered in the last Section.

It is well known that Weyl  points have  topological properties,~\cite{vol,exis,geo}  and  no  fine-tunning or symmetries are usually necessary for their existence. But  symmetry, while not necessary,   can sometimes  be  \emph{sufficient} for the existence of Weyl points. That is what we show in this paper, where we investigate the role played by   the space groups of  crystals with time reversal symmetry (TRS) in the existence of orbital Weyl points. 


The main results of this paper are summarized in Tables~\ref{t1}-\ref{t2}. Only crystals with one of the $19$ space groups in these tables \textit{can} have orbital Weyl points at  points of symmetry. Moreover, crystals with space groups in Table~\ref{t1} \textit{must} have orbital  Weyl points at  the listed  points,  \textit{irrespective of the actual crystal structure}. For crystals with space groups in  Table~\ref{t2} the situation is only slightly different: At the listed  points  both orbital  Weyl points and non-degenerate bands with quadratic dispersion relations are possible.
These results are relevant to nonmagnetic $3D$ crystals with sufficiently weak spin-orbit interactions and to cold atoms in optical lattices.   

The rest of the paper is organized as follows.  Our main results, contained in Eq.~(\ref{DW}) and  Tables~\ref{t1}-\ref{t2} are explained in Section~II. Section~III considers the effects of spin-orbit interactions on the orbital Weyl points, and a simple tight-binding model is constructed and analyzed in Section~IV. Possible extensions of our approach to Weyl semimetals, crystals with magnetic order and  other  types of Dirac points are considered in Section~V.
An outline of the methods used to obtain Tables~\ref{t1}-\ref{t2} is given in the Appendix.

\section{Orbital Weyl points and crystal symmetry}

Our strategy is based on  the fact that the form of the hamiltonian in the vicinity of a point of symmetry $\vec K_1$  is strongly constrained by the symmetries of the point in question.~\cite{slon,elph,trig,ek}  These include $\mathcal{G}_{K_1}$ ---the little group~\cite{liu,sym} of the vector $\vec K_1$---  and  combinations of  TRS with space group elements. Since we are interested in systems with very weak spin-orbit interactions, we will consider first the structure of the orbital or spin-independent part of the hamiltonian. The transformation properties of orbital wavefunctions are described by single-valued~\cite{brad}  representations of the space group. 

As explained in the Appendix, we have carried out a survey of all the single-valued irreducible representations of the $230$ space groups at  points of symmetry in the Brillouin zone (BZ). We find that, for most  space groups, the constraints on the form of the hamiltonian around points  of symmetry  are incompatible  with the structure of the  Weyl hamiltonian.  The comparatively few  exceptions  are listed in Tables~\ref{t1}-\ref{t2}. In all  cases, two electronic bands transforming according to a single-valued irreducible representation (IR) of $\mathcal{G}_{K_1}$ are degenerate at the point of symmetry $\vec K_1$.  Near the point of symmetry, i.e., for $\vec K=\vec K_1+\vec k$,  the degeneracy is broken by $\vec k$-dependent terms  and the hamiltonian takes the form
\beq\label{ham1}
H(\vec k)=v_x \s_x k_x+v_y \s_y k_y+v_z \s_z k_z+O(k^2)
\eeq
where $v_x=v_y$ for uniaxial crystals and $v_x=v_y=v_z$ for cubic crystals. After appropriate rescalings of the components of $\vec k$ for non-isotropic crystals, this is just the  
Weyl hamiltonian $H\sim v\vec \s\cdot  \vec k$, with the sign of $v$ equal to the product of the signs of $v_i$. The  points of symmetry and IRs where this happens are listed in the last column of Tables~\ref{t1}-\ref{t2} in standard notation.~\cite{brad}

TRS reverses the sign of $\vec K$. In those cases where  $-\vec K_1$ is not equivalent to $\vec K_1$, we get a copy of the Weyl  hamiltonian at the mirror point $-\vec K_1$ and fermions have, in addition to the pseudospin index associated to the Pauli matrices $\s_i$, a  `valley' index. As  shown in the Appendix, the total hamiltonian is then given by the  $4\times 4$ matrix 
\beq\label{DW}
H(\vec k)=v \left(
\begin{array}{cc}
 \vec \s\cdot  \vec k & 0\\
 0 &   \vec \s\cdot  \vec k \\
 \end{array}
\right)+O(k^2)
\eeq 
This describes two degenerate massless fermions of the \emph{same chirality}, right-handed for $v>0$, left-handed otherwise. Somewhat surprisingly, we find that this doubling continues to take place even  when 
 $-\vec K_1\equiv \vec K_1$, i.e., for TRS invariant momenta. In that case, Eq.~(\ref{DW}) describes two distinct fermions of the same chirality at the  \emph{same point} in the BZ. There is still a `valley'  index but, unlike in graphene, it can not be associated with two different points in the BZ. This happens  for the six space groups with simple (P) Bravais lattices in Table~\ref{t1}.
 
The groups in Table~\ref{t1} have one important feature in common: The IRs in the last column of the table include \emph{all} the IRs at the point of symmetry. This means that, at that point, all the bands must form  degenerate pairs with  Weyl hamiltonians. In other words, \emph{any} crystal with space group in Table~\ref{t1} will have bulk chiral fermions described by Eq.~(\ref{DW}), irrespective of the concrete crystal structure. 
 On the other hand, the space groups in Table~\ref{t2} have, besides  the listed small IRs $K_3$ and $H_3$, which are two-dimensional and give rise to orbital Weyl points, other one-dimensional  small IRs ($K_1,K_2,H_1,H_2$) not related  to Weyl points.   In this case,  both orbital Weyl points and non-degenerate bands with quadratic dispersion relations are possible at  the listed points of symmetry.

 \begin{table}[t]
\begin{tabular}{l l  l  l   }
 Space &Group   &  \;\;\;\;\;\; \;\;\; \;\;\; \;\;\; &IRs\\
\hline \hline
$\mathbf{214}$ & $I4_132\;\;\;\;\;$ & $O^8\;\;\;\;\;\; \;\;\; \;\;\;$ & $P_1,P_2,P_3$\\
$\mathbf{213}$ & $P4_132$ & $O^7$ &$R_1,R_2,R_3$\\
$\mathbf{212}$ & $P4_332$ & $O^6$ &$R_1,R_2,R_3$\\
$\mathbf{199}$ & $I2_13$ & $T^5$ & $P_1,P_2,P_3$\\
$\mathbf{198}$ & $P2_13$ & $T^4$ &$R_1,R_2,R_3$\\
$\mathbf{098}$ & $I4_122$ & $D_4^{10}$ &$P_1$\\
$\mathbf{096}$ & $P4_32_12$ & $D_4^{8}$ &$A_1,A_2$\\
$\mathbf{092}$ & $P4_12_12$ & $D_4^{4}$ &$A_1,A_2$\\
$\mathbf{024}$ & $I2_12_12_1$ & $D_2^{9}$ &$W_1$\\
$\mathbf{019}$ & $P2_12_12_1$ & $D_2^{4}$ &$R_1$\\
\end{tabular}
\caption{ Cubic, tetragonal and orthorhombic space groups with orbital Weyl points. The small IRs are all $2d$ (except for $R_3$ of $\mathbf{212}$ and $\mathbf{213}$, which is $4d$) and refer to the symmetry points $P(\qua,\qua,\qua)$, $R(\med,\med,\med)$, $A(\med,\med,\med)$ and $W(\frac{3}{4},\frac{\bar 1}{4},\frac{\bar 1}{4})$, with components in the conventional basis of Ref.~\cite{brad}  The stars have two vectors $(\vec K_1,-\vec K_1)$  for body-centered ($I$) lattices, and a single vector $\vec K_1\equiv-\vec K_1$ for  simple ($P$) lattices.}
\label{t1}
\end{table}

 \begin{table}[b]
\begin{tabular}{l l  l  l   }
 Space &Group   &  \;\;\;\;\;\; \;\;\; \;\;\; \;\;\; &IRs\\
\hline \hline
$\mathbf{182}$ & $P6_322\;\;\;\;\;$ & $D_6^6\;\;\;\;\;\; \;\;\; \;\;\;$ & $K_3$\\
$\mathbf{181}$ & $P6_422\;\;\;\;\;$ & $D_6^5\;\;\;\;\;\; \;\;\; \;\;\;$ & $K_3,H_3$\\
$\mathbf{180}$ & $P6_222\;\;\;\;\;$ & $D_6^4\;\;\;\;\;\; \;\;\; \;\;\;$ & $K_3,H_3$\\
$\mathbf{179}$ & $P6_522\;\;\;\;\;$ & $D_6^3\;\;\;\;\;\; \;\;\; \;\;\;$ & $K_3$\\
$\mathbf{178}$ & $P6_122\;\;\;\;\;$ & $D_6^2\;\;\;\;\;\; \;\;\; \;\;\;$ & $K_3$\\
$\mathbf{177}$ & $P622\;\;\;\;\;$ & $D_6^1\;\;\;\;\;\; \;\;\; \;\;\;$ & $K_3,H_3$\\
$\mathbf{154}$ & $P3_221\;\;\;\;\;$ & $D_3^6\;\;\;\;\;\; \;\;\; \;\;\;$ & $K_3,H_3$\\
$\mathbf{152}$ & $P3_121\;\;\;\;\;$ & $D_3^4\;\;\;\;\;\; \;\;\; \;\;\;$ & $K_3,H_3$\\
$\mathbf{150}$ & $P321\;\;\;\;\;$ & $D_3^2\;\;\;\;\;\; \;\;\; \;\;\;$ & $K_3,H_3$\\
\end{tabular}
\caption{ Hexagonal and trigonal space groups with orbital Weyl points. The small IRs listed are all $2d$ and refer to the symmetry points $K(\frac{\bar 1}{3}, \frac{2}{3},0)$ and $H(\frac{\bar 1}{3}, \frac{2}{3}, \med)$, with components in the conventional basis of Ref.~\cite{brad} The stars have two vectors $(\vec K_1,-\vec K_1)$ in all cases.}
\label{t2}
\end{table}

 We close this section by pointing out some special features in Tables~\ref{t1}-\ref{t2}. 
The first one is that all the groups in Table~\ref{t1} are subgroups of the first entry,~\cite{mois1,mois2} the  cubic  space group $\mathbf{214}\;I4_132$ and, despite the use of different conventional names ($P,R,A,W$), they also share the point of symmetry. Indeed,   the cartesian coordinates for all the  points in Table~\ref{t1} can be written as
\beq\label{point}
\vec K_1=(\frac{\pi}{a}, \frac{\pi}{b}, \frac{\pi}{c})
\eeq
in terms of the unit cell constants, with $b=c$ for uniaxial crystals and $a=b=c$ for cubic crystals. 
As a result, one can begin with any $3D$ lattice with space group $\mathbf{214}$ and reproduce all the cases in Table~\ref{t1} by suitable deformations. The second somewhat surprising feature is that all the stars have just one or two vectors. As a consequence, these crystals have only one or two  orbital Weyl points degenerate in energy. This should be contrasted, for instance, with the case studied in Ref.,~\cite{arc} where $24$ Weyl points (away from points of symmetry) are present at the Fermi energy.

\section{Spin-orbit interactions}

\begin{figure}[t]
\begin{center}
\includegraphics[angle=0,width=0.85\linewidth]{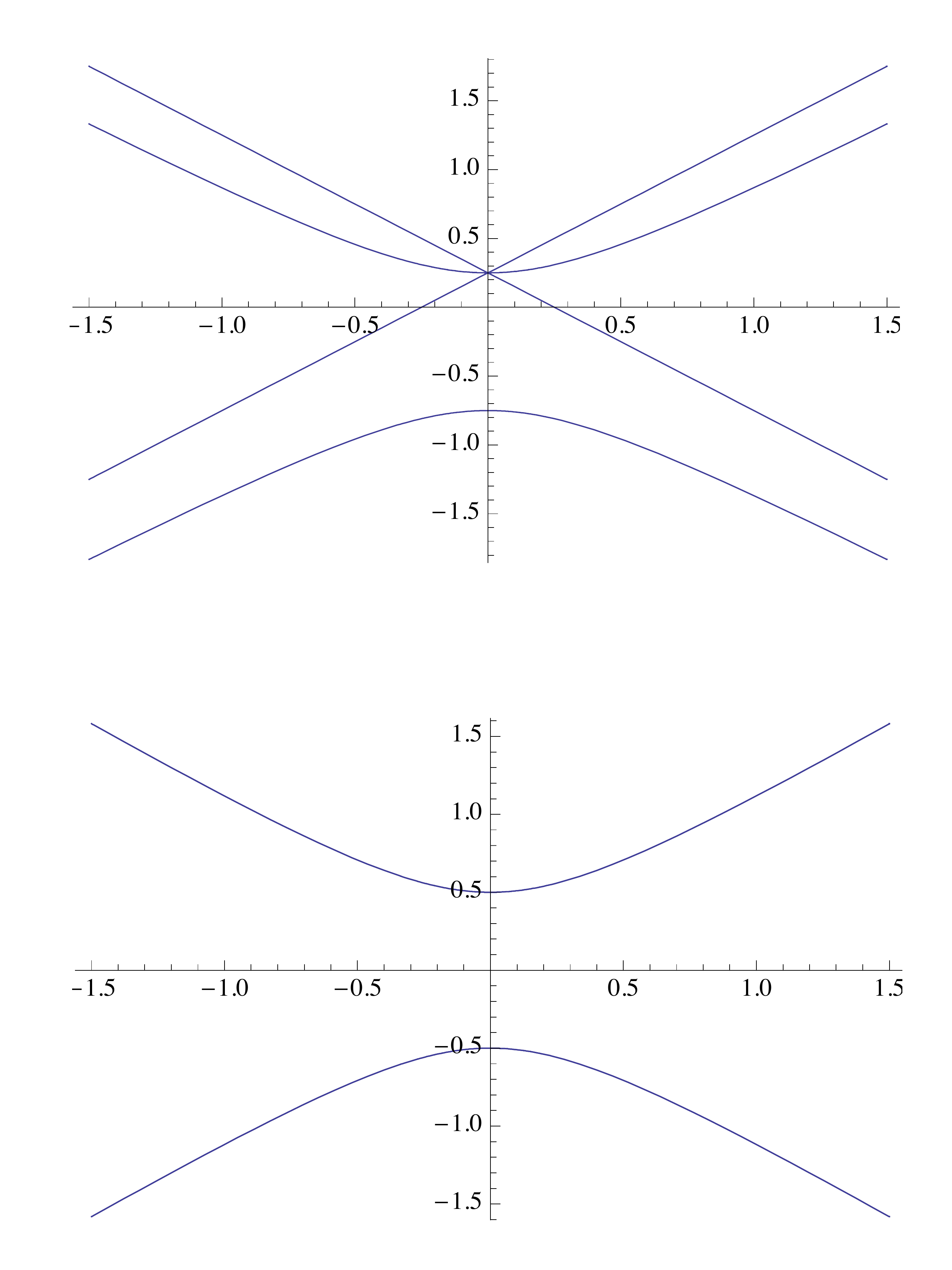}
\end{center}
\caption{ Effects of spin-orbit coupling on the orbital Weyl point of a   cubic crystal (top) and graphene (bottom) in arbitrary units.}
\label{f3}
\end{figure}

Strictly speaking, our analysis so far  applies only  to `spinless electrons'. 
It is well known that spin-orbit interactions in graphene open a gap and give fermionic excitations a small mass.~\cite{km}  In the case of graphene, 
 the intrinsic spin-orbit hamiltonian is proportional  to $\sigma_z\otimes  s_z\otimes\tau_z$, where $s_z$ and $\tau_z$ are Pauli matrices for electron spin and valley indices respectively. Around each valley, the hamiltonian can be written in terms of $4\times 4$ matrices
 \beq\label{dirac4}
 H_0+H_{so}\!=\! v(\a_x k_x+\a_y k_y)+\beta\Delta
 \eeq
 where $\a_i\!=\!\s_i\otimes\mathbf{1}_s$, $\beta\!=\!\pm\s_z\otimes s_z$ and $\Delta$ is the strength of the spin-orbit coupling. The matrices satisfy the Clifford algebra 
 \beq\label{clif}
 \{\a_i,\a_j\}\!=\!2\delta_{ij}\;\!,\;\; \{\a_i,\beta\}\!=\!0\;\!,\;\;\beta^2\!=\!2
 \eeq
 This identifies Eq.~(\ref{dirac4}) as the Dirac hamiltonian for \hbox{$4$-component}  fermions with mass $m\!=\!\Delta$ and spectrum $E_{\pm}\!=\!\pm\sqrt{\Delta^2+v^2  k^2}$.

For the space groups in Tables~\ref{t1}-\ref{t2}, the most general \hbox{$k$-independent} spin-orbit hamiltonian compatible with  spatial symmetries  and TRS takes the form
\beq\label{hso}
H_{so}\!=\!\frac{1}{4} \left(\Delta_x\sigma_x\otimes s_x\!+\!\Delta_y\sigma_y\otimes s_y\!+\!\Delta_z\sigma_z\otimes s_z\right)\otimes\mathbf{1}_\tau
\eeq
where $\Delta_x=\Delta_y$ for uniaxial crystals and \hbox{$\Delta_x=\Delta_y=\Delta_z$} for cubic crystals.  
Actually, this valley independent  form of the spin-orbit interaction is valid on the basis $(e_1, e_2, i e_2^*, -i e_1^*)$ of  orbital wavefunctions introduced in the Appendix.  If one uses the more conventional basis $(e_1, e_2,  e_1^*,  e_2^*)$, then   one has to append  the valley matrix $\tau_z$ to the $x$ and $z$ components in Eq.~(\ref{hso}).  Note that, in the conventional basis, instead of Eq.~(\ref{DW}), we would have Eq.~(\ref{DW2}).

The spectrum of the total hamiltonian $v\vec\sigma\cdot \vec k+H_{so}$ can be computed numerically and one finds that, in general, gaps are generated and all fermionic excitations acquire masses. Cubic crystals, where $H_{so}$ is isotropic and depends on a single parameter $\Delta$,  are an  exception and can be treated analytically. In this case the spectrum is given by 
\beqa
E^-_{\pm}&=&-\frac{\Delta}{4}\pm\sqrt{(\frac{\Delta}{2})^2+v^2 k^2}\non\\
E^+_{\pm}&=&\frac{\Delta}{4}\pm v  k
\eeqa
and contains  massless excitations. This spectrum is radically different from that of the Dirac hamiltonian appropriate for   $2D$  graphene. Note, in particular,  that the linear bands $E^+_{\pm}$, together with  $E^-_+$, form a triplet  (see Fig.~\ref{f3}), following the usual rules for addition of angular momenta with $L\!=\!S\!=\!1/2$ and $\vec J\!=\!\vec L+\vec S$. This is only natural: assembling  pseudospin and spin into a $4$-component object is equivalent to taking the Kronecker product of two $j=1/2$ irreducible representations of the $SO(3)$ rotation group and this product decomposes according the rules of angular momentum composition.~\cite{brad}

Thus, unlike in 2D where strong spin-orbit interactions simply destroy the orbital Weyl points and turn massless  fermions  into massive excitations, here  we also  get Dirac points with $J\!=\!1$. In the next Section we will present a model that in the absence of spin-orbit   has, besides  orbital Weyl points,  Dirac points with pseudospin-one. The points with pseudospin-one would be split by strong spin-orbit  interactions  into Weyl points with $J=1/2$ and novel Dirac points with $J=3/2$.  Subduction of IRs~\cite{sym,brad} can be used in principle to extend the analysis to non-cubic groups. 

Henceforth we will assume that spin-orbit interactions are weak and can be ignored. In this limit the dynamics is well described by the $3D$ Weyl hamiltonian --albeit with an additional two-fold degeneracy due to spin-- and, as a consequence, the systems considered in this paper may share some of the properties of Weyl semimetals.

\section{A tight-binding example}

As a practical application, we  present a  tight-binding model with space group $\mathbf{214} \;I4_132 (O^8)$, the first entry in Table~\ref{t1}. According to our previous analysis, \emph{any} lattice with this space group must  have orbital Weyl points at $P$. Here we consider a lattice with  four atoms per primitive unit cell, with cartesian coordinates \hbox{$\vec r_1=a/8(1,1,1)$}, $\vec r_2=a/8(1,\bar 1,3)$, $\vec r_3=a/8(3,\bar 1,5)$ and $\vec r_4=a/8(3, 1,7)$.~\footnote{We are using  Wyckoff position~\cite{mois1,mois2} $8a$ because of its relatively low multiplicity.} 

\begin{figure}[t]
\begin{center}
\includegraphics[angle=0,width=0.85\linewidth]{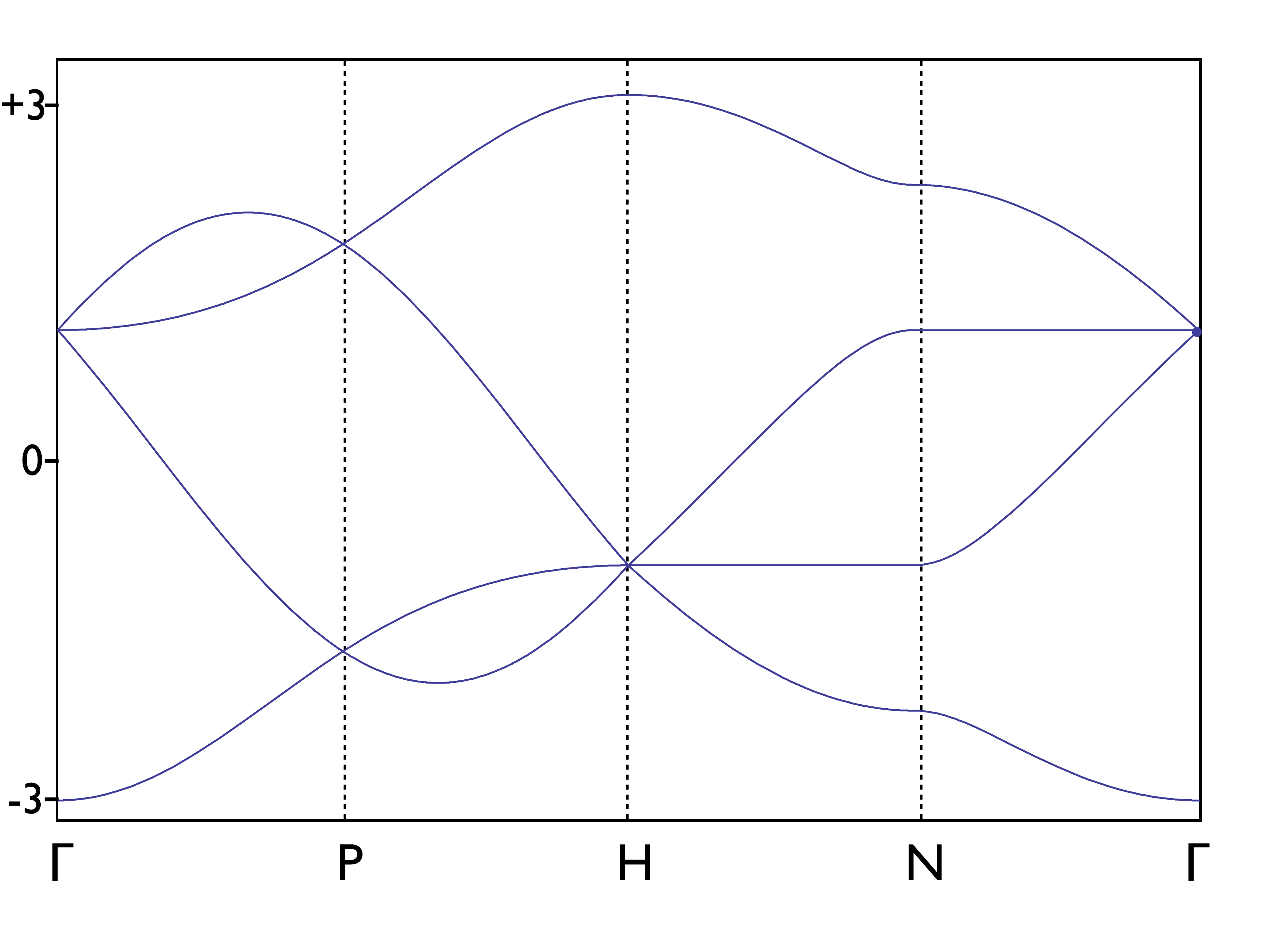}
\end{center}
\caption{ Bands for the cubic lattice in the NN approximation for $t=-1$. The  BCC Brillouin zone with its points and lines of symmetry can be seen in Fig.~\ref{f4}.}
\label{f1}
\end{figure}

To each atom we associate a Bloch function
\beq\label{bloch}
\Phi_{i} (\vec k)=\sum_{\vec t \epsilon \mathcal{T}} e^{i \vec k\cdot(\vec r_i+ \vec t)} \varphi (\vec r-\vec r_i-\vec t) 
\eeq
where the sum runs over all the points in the Bravais lattice and $\varphi(\vec r)$ is an $s$-wave atomic orbital. Each atom has three nearest neighbors (NN), with bonds parallel to the three cartesian planes. In terms of the reduced cartesian components of the wave vector $\vec k=2\pi/a(k_x,k_y,k_z)$  the NN tight-binding hamiltonian is given by 
\begin{widetext}
\beq\label{lcao4}
H(\vec k)=t\left(
\begin{array}{cccc}
 0 & e^{\frac{i\pi}{2}    (k_z-k_y)} &
    e^{\frac{i\pi}{2}  (k_y-k_x)}  &
   e^{\frac{i\pi}{2} (k_x-k_z)} \\
 e^{-\frac{i\pi}{2}   (k_z-k_y)} & 0
   &   e^{\frac{i\pi}{2}  (k_x+k_z)} &
   e^{-\frac{i\pi}{2}  (k_x+k_y)} 
   \\
  e^{-\frac{i\pi}{2}  (k_y-k_x)}  &
   e^{-\frac{i\pi}{2}   (k_x+k_z)} &
   0 & e^{\frac{i\pi}{2}   (k_y+k_z)}
   \\
 e^{-\frac{i\pi}{2}   (k_x-k_z)} &
    e^{\frac{i\pi}{2}   (k_x+k_y)}  &
   e^{-\frac{i\pi}{2}  (k_y+k_z)} &
   0
\end{array}
\!\!\!\right)
\eeq 
\end{widetext}
where $t<0$ is the hopping parameter and the diagonal on-site energy has been set to zero. The hamiltonian can be diagonalized numerically and the resulting bands are shown in Fig.~\ref{f1}. 

\begin{figure}[b]
\begin{center}
\includegraphics[angle=0,width=0.95\linewidth]{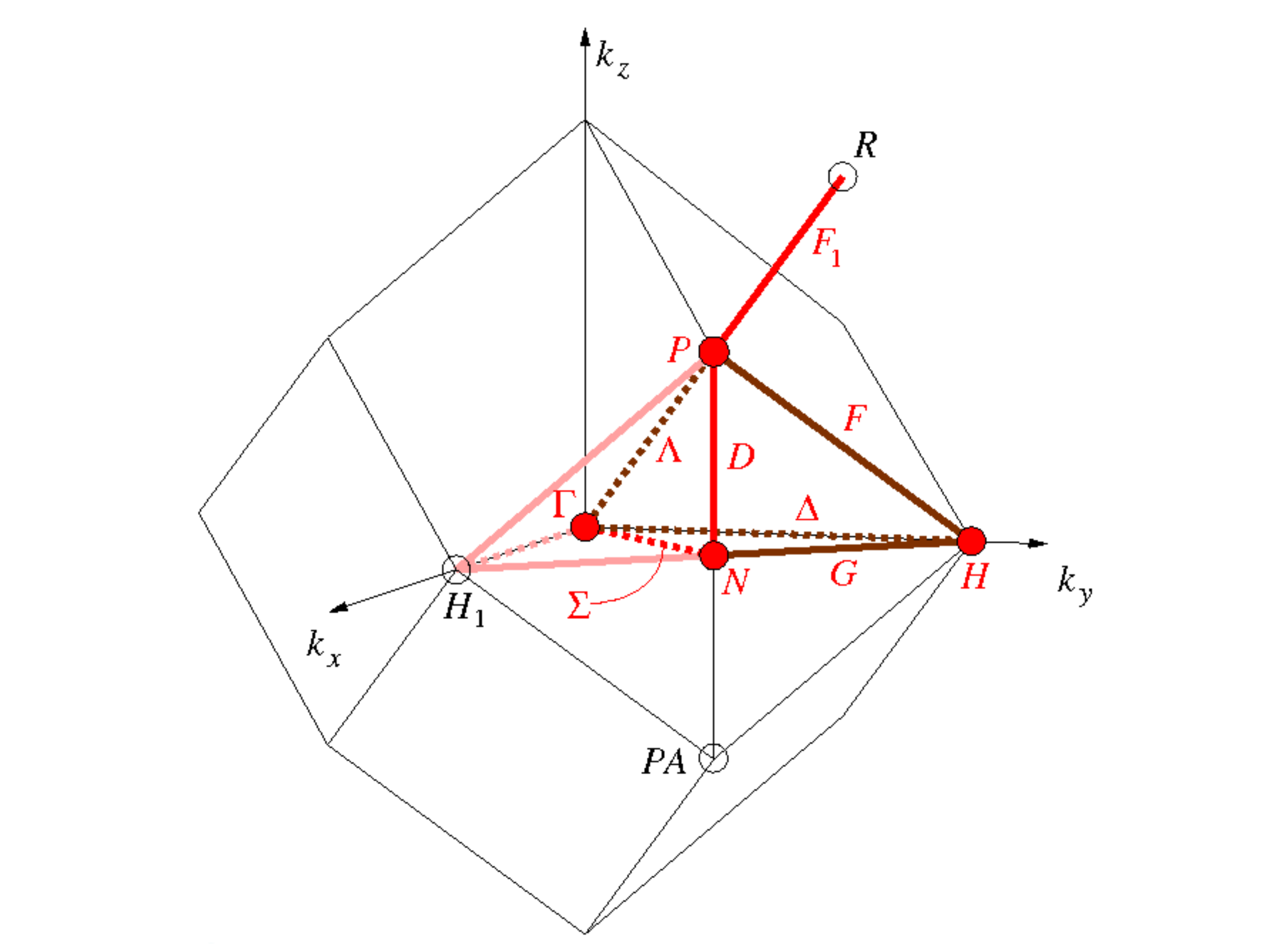}
\end{center}
\caption{Brillouin zone for BCC crystals.~\cite{mois1,mois2}}
\label{f4}
\end{figure}
 
 The existence of linear bands at point $P$ is obvious in Fig.~\ref{f1}. We can use standard group theory techniques to confirm that they  actually are Weyl points, as  predicted. The four Bloch functions $\Phi_i$ form the basis of a reducible representation $\mathcal{H}$ that can be decomposed into small IRs as $\mathcal{H}(\Phi_1,\ldots,\Phi_4)=P_2(e_1,e_2)+P_3(u_1,u_2)$. Up to normalizations, the symmetry-adapted vectors are given by $e_1\sim(1,\beta_2,-1,i \beta_2)$, $e_2\sim(-\beta_2,-1,-\beta_2,i)$, with $\beta_2=(1+\sqrt{3})(1-i)/2$, and identical expressions for $u_1,u_2$ with $\beta_2$ replaced by $\beta_3=(1-\sqrt{3})(1-i)/2$. Using a unitary transformation $U_P$ to change to the symmetry adapted basis and expanding around the point $P$ yields
\beq\label{hlin}
U_P^\dagger H(\vec k) U_P= \frac{ \pi t}{\sqrt{3}}\left(
\begin{array}{cc}
 \frac{3}{\pi}+\vec\sigma\cdot\vec k & H_{23}(\vec k)  \\
 H^\dagger_{23}(\vec k) &  -\frac{3}{\pi}-\vec\sigma\cdot\vec k \\
 \end{array}
\right)+O(k^2)
\eeq 
where $H_{23}(\vec k)$ is given by
\beq\label{h23}
H_{23}(\vec k)= \frac{ 1}{\sqrt{2}}\left(
\begin{array}{cc}
k_z & \om^* k_x-i\om k_y\\
 \om^* k_x+i\om k_y &  -k_z  \\
 \end{array}
\right)
\eeq 
with $\om=e^{2\pi i/3}$. There are thus two orbital Weyl points at $P$ with different energies $\pm t\sqrt{3}$ and opposite chiralities. The existence of orbital Weyl points of opposite chirality   is of course to be expected from fermion doubling,~\cite{nimo} which requires the net chirality of the BZ to vanish, although the fact that they appear at coincident points is peculiar to this model.  Note also the  linear couplings between the two points. Due to the split in energies, these couplings contribute  corrections $O(k^2)$ to the $2\times 2$ effective hamiltonians around the orbital Weyl points and do not spoil their structure.
A similar expansion around $-\vec K_1$ confirms  that, due to TRS,  each orbital Weyl point is degenerate in energy with another point of the same chirality. Restoring the lattice constant yields a  Fermi velocity  $v_F=\frac{a|t|}{2\sqrt{3}}$. Note however that, as no symmetry connects the orbital Weyl points with different chiralities and energies,  going beyond the NN approximation  is expected to give different Fermi velocities for them.

Fig.~\ref{f1} exhibits  linear bands around the $\Gamma$ and $H$ points as well. Their nature is, however, very different from that of the  orbital Weyl points at $P$. Let's consider, for the sake of concreteness, the  $\Gamma$ point. In this case, the representation associated with the Bloch functions decompose into IRs of dimension one  and three,\hbox{ $\mathcal{H}=A_1+T_2$}. Transforming to the appropriate   symmetry-adapted basis and linearizing yields 
\beq\label{hling}
U_\Gamma^\dagger H(\vec k) U_\Gamma=t\left(
\begin{array}{cccc}
 3 &0 & 0 & 0 \\
0 & -1 &   -i \pi k_z& i \pi k_y \\
0 & i \pi k_z &   -1 &- i \pi k_x\\
0 & - i \pi k_y &   i \pi k_x& -1 \\
 \end{array}
\right)+O(k^2)
\eeq
Up to a constant energy, the $3\times 3$ block can be written
\beq\label{spin1}
H_{T_2}(\vec k)=\pi t\vec J\cdot \vec k
\eeq
where $(J_i)_{jk}=-i \varepsilon_{ijk}$ are spin-$1$ matrices. Thus, around this Dirac   point, electrons behave more like massless spin-\emph{one} particles, with spectrum $E(\vec k)=0, \pm v_F |\vec k|$ and  $v_F=\med a |t|$. Indeed, one can check that, while  the \hbox{$E=0$} component is longitudinally polarized,  the other two are transverse, just like the propagating components of a photon. Pseudospin-one Dirac points have been  reported in some two~\cite{one1,one2,one3} and three-dimensional~\cite{one4} systems.

\begin{figure}[t]
\begin{center}
\includegraphics[angle=0,width=0.85\linewidth]{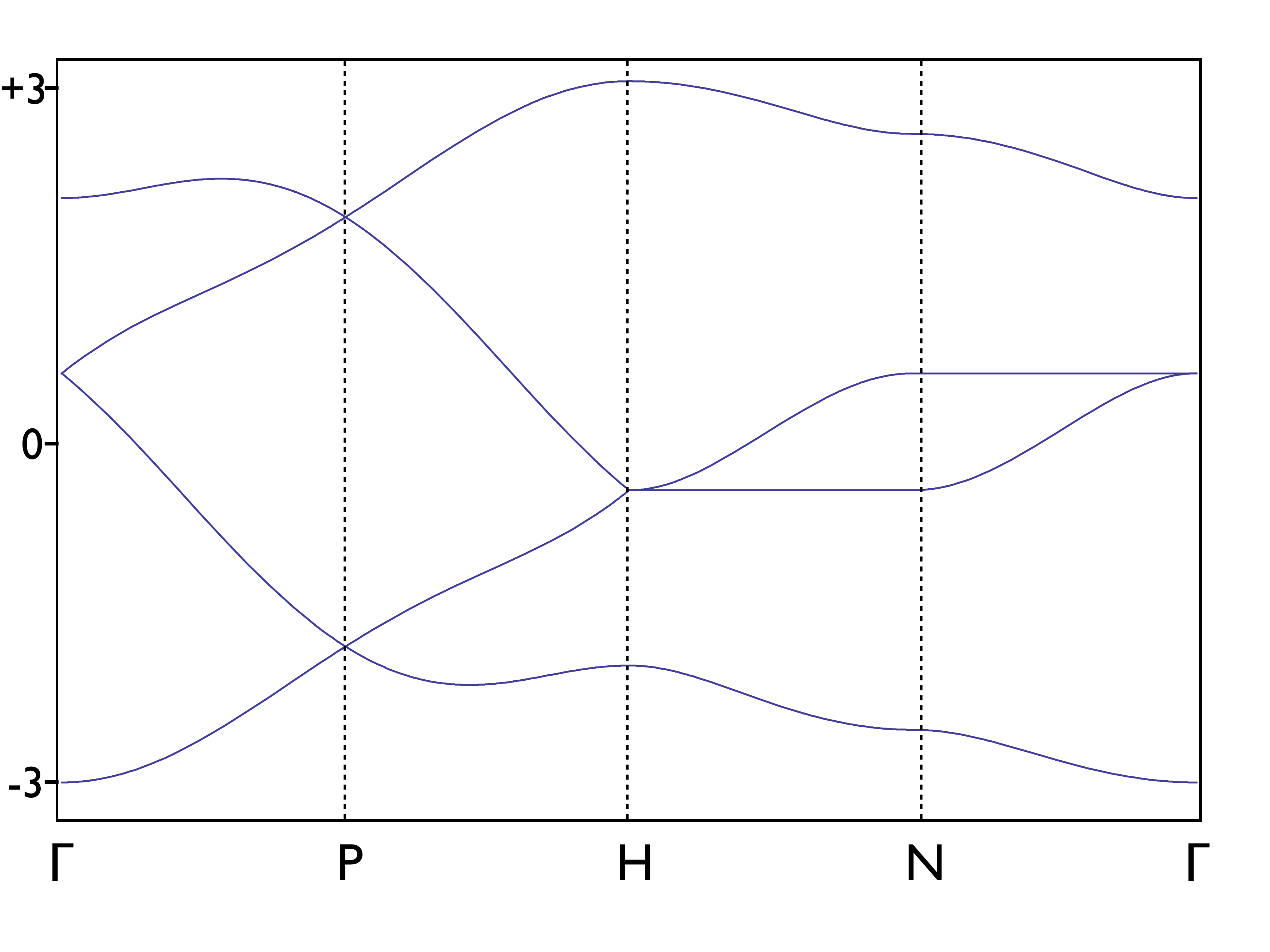}
\end{center}
\caption{ Bands for the cubic lattice with a tetragonal distortion ($\epsilon=0.4$) for $t=-1.25$. See the main text for details. }
\label{f2}
\end{figure}

A look at  Fig.~\ref{f1} shows that,  even if the Fermi level coincides with one of the orbital Weyl points at $P$, band overlap will cause the dynamics to be dominated by large electron (or hole) pockets. The existence of band overlap can be traced in this case to the  $3$-fold degeneracies  in Fig.~\ref{f1}, which force one of the bands arising from the orbital Weyl points to bend over. As $3d$ IRs exist only for cubic groups, we may modify the model by making 
the hopping parameters $t_\perp$ associated with bonds parallel to the $OXY$-plane  different from the rest, $t_\perp=\epsilon\, t$. This reduces the symmetry to the tetragonal subgroup $\mathbf{98}\; I4_122$ and eliminates the $3$-fold degeneracies at $\Gamma$ and $H$.
Fig.~\ref{f2} shows the bands for $\epsilon=0.4$. In the absence of spin-orbit interactions  the system would behave like 
  a gapless semiconductor with massless carriers for $1/4$ and $3/4$  fillings, with  positive or negative chirality depending on the filling.

\section{Discussion}

In this paper we  have studied the interplay between crystal symmetry and the existence of bulk chiral fermions in nonmagnetic  $3D$ crystals with weak spin-orbit coupling. We have shown  that the space group plays a  determinant role, as summarized in Tables~\ref{t1}-\ref{t2}. As Weyl semimetals depend on the existence of magnetic order or  strong spin-orbit interactions,~\cite{trans}  we have explored  an entirely different corner in the space of candidate $3D$ crystals with bulk chiral fermions.  
We have  also  considered the effects of spin-orbit interactions and shown that, unlike in $2D$,  these may give rise to new critical points supporting massless fermions for different values of the pseudospin. 
For sufficiently weak spin-orbit interactions the dynamics  is well described by the $3D$ Weyl hamiltonian and the systems considered in this paper may share some of the properties of Weyl semimetals.

There are  two obvious uses for  the information in Tables~\ref{t1}-\ref{t2}. First, one can  look for orbital Weyl points in  nonmagnetic crystals with weak spin-orbit interactions and  space groups in the tables, either in  theoretically computed  electronic bands or experimentally. 
The other use is the design of $3D$ lattices with Weyl points. This may allow for a physical realization of massless chiral fermions with cold atoms in optical lattices. 
Note  that the physical relevance as well as the feasibility of detecting  chiral fermions in actual crystals will be affected by  structure dependent features such as the position of the Fermi level and  the amount of band overlap. But knowing that the bulk chiral  fermions \emph{have} to be there is obviously a good starting point. As shown in the example of Section~IV,  we can then try to engineer the required properties  by modifying  the initial system.

There are several possible extensions to this work. One is to consider the existence of orbital Weyl points away from points of symmetry. All the points in a \emph{line} of symmetry, except for the endpoints, have the same group $\mathcal{G}_{K_1}$. Therefore, symmetry alone can not imply the existence of bulk chiral fermions at a particular point in the line, although it will indicate whether this is possible at all.  But it can, in some cases, imply the existence along the line of `semi-Dirac' points, where the dispersion relations are linear only for some directions. For generic points in the BZ, symmetry alone has little to say and a different kind of analysis may be useful.~\cite{design,dlat}

Sometimes not two, but three bands become degenerate at a Dirac point.\cite{one4} This is the case with  the $\Gamma$ and $H$ points in the example of Section~IV. This is possible for some cubic space groups, and group theory can be used to determine the IRs and points of symmetry where this may happen. As shown in Section~III, turning on spin-orbit interactions will give rise to novel Dirac points with pseudospin $3/2$. Other, more complicate linear hamiltonians\cite{graphi,al,fine} are also possible and may be physically relevant. Group theory can be used to classify or even  predict the existence of the different types of Dirac points.
 
In this paper we have analyzed all the single-valued irreducible representations  at points of symmetry in the BZ. These are appropriate for orbital degrees of freedom. By considering instead 
double-valued~\cite{brad} IRs we could study the existence  of Weyl points where the total hamiltonian, including electron spin and spin-orbit interactions, takes the form of the $3D$ Weyl hamiltonian. Thus, we could extend our analysis to TRS invariant Weyl semimetals. Double-valued IRs have been recently applied to the study  of `Dirac semimetals'.~\cite{dsm}
  
So far we have restricted ourselves to TRS invariant crystals. The reasons  are mostly practical. The symmetries of crystals with magnetic order are classified by the $1651$ magnetic space groups~\cite{brad}, instead of the $230$ ordinary (Fedorov) space groups that classify crystals with TRS. The amount of work required to examine the points of symmetry and irreducible corepresentations~\cite{brad} for  all  the magnetic groups  is, obviously, much greater. Moreover, unlike  ordinary space groups, there are few databases with the magnetic space groups of crystals with magnetic order. 

Nevertheless, some of the results in this paper can also be applied  to spinless electrons in crystals with magnetic order. The reason is that, by construction,  the $19$ entries in Tables~\ref{t1}-\ref{t2}  describe situations where the $3D$ Weyl hamiltonian is invariant under the `grey' or `type II' Shubnikov space group~\cite{brad} associated to  an ordinary (Fedorov) space group by the addition of the TRS operation. Now, all the magnetic space groups derived from the Fedorov space group
with the BNS settings~\cite{mois1,mois2} are subgroups of the grey group. As a consequence, the corresponding Weyl hamiltonian is automatically invariant under \emph{all} the magnetic groups derived from the ordinary space groups listed in our tables. For instance, the Weyl hamiltonians at the $P$ point of group $\bf{214}$ are automatically invariant under the derived magnetic groups $\mathbf{214.68}$ and $\mathbf{214.69}$ (in the BNS settings). 

One still has to check  that the degeneracy of the two bands at the point of symmetry, necessary for the existence of the Weyl point, is maintained under the lower symmetry of the magnetic space group. If this is not the case but the effects of the magnetic order are small,  the Weyl point will survive, but  move away from the point of symmetry. 
The main difference when dealing with  crystals with magnetic order is that we can not exclude the possibility  of finding  bulk  chiral fermions around points of symmetry  for space groups \emph{not} listed in Tables~\ref{t1}-\ref{t2}. That happens whenever the Weyl hamiltonian is invariant under the magnetic subgroup of the grey space group, but not under the grey space group itself. 

\section*{Acknowledgments}
It is a pleasure to thank  F. Guinea  for useful comments on an earlier draft of this paper and to J. M. P\'erez-Mato for help in using the Bilbao Crystallographic Server~\cite{mois1,mois2}(http://www.cryst.ehu.es). 
This work is supported in part by the Spanish Ministry of Science and Technology under Grant FPA2009-10612 and  the Spanish Consolider-Ingenio 2010 Programme CPAN (CSD2007-00042), and by the Basque Government under Grant IT559-10.

\appendix
\section{}
In this Appendix we  outline the methods used to obtain Tables~\ref{t1}-\ref{t2} and Eq.~(\ref{DW}).
 Symmetry operations belonging to the space group $\mathcal{G}$ will be written $g=\{\a |\vec v\}$, where $\a$ and $\vec v$ denote the rotation and translation parts respectively.~\cite{brad} 
Let $\{ e_i (\vec K)\}$ be a basis of orbital wave functions that transform linearly under the action of $\mathcal{G}$, $e_i (\vec K) \to e_j(\a\vec K) R_{ji}(g)$, 
where we sum over repeated indices. Invariance of the hamiltonian under $\mathcal{G}$ means that, for any wavefunction $\psi$, $\langle H\rangle_\psi=\langle H\rangle_{\psi_g}$, where ${\psi_g}$ is the transformed of $\psi$ by the group element $g$. Expanding this condition on the  basis $\{ e_i (\vec K)\}$ yields, in matrix notation 
\beq\label{invuni}
R^\dagger(g) H(\a\vec K) R(g)= H(\vec K)
\eeq
where $H_{ij}(\vec K)=\langle e_i (\vec K)|H| e_j(\vec K)\rangle$. 
Time reversal $\theta$ is an antiunitary operation that  acts on  orbital wavefunctions by complex conjugation, $e_i (\vec K) \to e_i (\vec K)^*= e_j(-\vec K) \Theta_{ji}$, 
where $\Theta_{ij}$ is a \emph{unitary} matrix,~\cite{liu,brad} and reverses the momentum $\vec K$. Invariance under $\theta$ implies
\beq\label{invtime}
\Theta^\dagger H(-\vec K) \Theta= H^*(\vec K)
\eeq
We will also have to consider combined antiunitary operations of the form $\theta g$. In this case invariance of the hamiltonian implies 
\beq\label{invtcomb}
T^\dagger(g) H(-\a\vec K)T(g)= H^*(\vec K)
\eeq
where $T(g)=\Theta R^*(g)$. Eqs.~(\ref{invuni},\ref{invtime},\ref{invtcomb})  become powerful constraints on the form of the hamiltonian when we  take \hbox{$\vec K=\vec K_1+\vec k$} in the neighborhood of a point of symmetry $\vec K_1$ and consider a power expansion in $\vec k$. 

Consider for instance Eq.~(\ref{invuni}) with $g$ restricted  to the little group of  $\vec K_1$, $ \mathcal{G}_{K_1}$, i.e., $\a\vec K_1\equiv \vec K_1$
\beq\label{invuni2}
R^\dagger(g) H(\vec K_1+\a\vec k) R(g)= H(\vec K_1+\vec k)
\eeq
By definition, the matrix $\a$ belongs to the vector representation $V$.~\cite{liu,sym}  If the basis functions $\{ e_i (\vec K)\}$ belong to the small IR $R$ of $\mathcal{G}_{K_1}$,  terms of order $n$ in  $\vec k$ in an expansion of the l.h.s. of Eq.~(\ref{invuni2}) will transform according to the product $R^*\times R\times [V]^n$, where $R^*$ is the complex conjugate of $R$ and $[V]^n$ denotes the  $n$-th symmetric power of the representation $V$.~\cite{liu,sym}  Then one can  use standard group theory techniques to determine, order by order in $\vec k$, the most general form of the hamiltonian compatible with the symmetries of the vector $\vec K_1$. 
 
 In particular, a necessary condition for the existence of the Weyl hamiltonian, which is linear in $\vec k$, is that   the vector representation $V$ is contained in the product $R^*\times R$ for  some $2d$ IR~\footnote{The $4d$ IR $R_3$ of $\mathbf{212}$ and  $\mathbf{213}$  is the exception. In this case, the two degenerate chiral fermions  belong to a single $4d$ IR.} $R$ of $\mathcal{G}_{K_1}$. 
 We have  checked  this condition on all the single-valued $2d$ and $4d$ IRs at the points of symmetry of the $230$ space groups to obtain a first list of candidates.~\footnote{We have also examined the `physically irreducible' $2d$ representations, i.e., pairs of conjugate $1d$  representations degenerate by TRS. These are never compatible with the Weyl hamiltonian.} Polar groups have been discarded from the outset, as they couple one component of the momentum to the unit matrix and the result is incompatible with the structure of the Weyl hamiltonian. The amount of work involved at this stage has been substantially  reduced thanks to  the use of `abstract groups' in Ref.~\cite{brad}  In essence, the abstract groups represent classes of isomorphic  little groups  $ \mathcal{G}_{K_1}$ and their number is much smaller than the number of little groups.

In the next step each candidate IR has been checked for invariance of the corresponding hamiltonian under TRS. This involves using Eq.~(\ref{invtime}) whenever  $-\vec K_1\equiv \vec K_1$ and Eq.~(\ref{invtcomb}) if $\vec K_1$ is not equivalent to $-\vec K_1$ but there exists a space group element $g=\{\a |\vec v\}\in \mathcal{G}$   such that $-\a\vec K_1\equiv \vec K_1$. 
The IRs that pass this last test are listed in the last column of Tables~\ref{t1}-\ref{t2}. These IRs have an important property: There is always a basis where the matrices $R(g)$ coincide, up to $g$-dependent phases, with the $j=1/2$ rotation matrices
\beq\label{jmed}
R_{1/2}(\hat n\phi)=\exp\left(-\frac{i\vec \s\cdot\hat n\;\phi}{2}\right)
\eeq
where $\phi$ is the angle around the unit vector $\hat n$.
 As a consequence,  we can  always transform to a basis where, up to appropriate rescalings of the components of $\vec k$ for non-isotropic crystals (see Eq.~(\ref{ham1})), the linear hamiltonian takes the standard Weyl form $H(\vec K_1+\vec k)\simeq v\, \vec \s\cdot\vec k$.  

When  $-\vec K_1$ is not equivalent to $\vec K_1$, the basis at these two points can be related by TRS. Choosing as basis at $-\vec K_1$   the complex conjugate of the one at $\vec K_1$, i.e., taking as $4d$ basis $(e_1, e_2,  e_1^*,  e_2^*)$, and using  Eq.~(\ref{invtime}) gives    
\hbox{$H(-\vec K_1+\vec k)\simeq-v\, \vec \s^*\cdot\vec k$}. The off-diagonal blocks between $\vec K_1$ and $-\vec K_1$ vanish by translation invariance, and  we have
\beq\label{DW2}
H(\vec k)=v \left(
\begin{array}{cc}
 \vec \s\cdot  \vec k & 0\\
 0 &   -\vec \s^*\cdot  \vec k \\
 \end{array}
\right)+O(k^2)
\eeq 
We can make the symmetry between $\vec K_1$ and $-\vec K_1$ obvious by using  the  $SU(2)$ transformation  \hbox{$\s_y \vec \s^* \s_y=-\vec \s$} to change the basis at $-\vec K_1$ so that the $4d$ basis is $(e_1, e_2, i e_2^*, -i e_1^*)$. On this basis the hamiltonian takes the form given in Eq.~(\ref{DW}).  

The case  $-\vec K_1\equiv \vec K_1$ is more subtle,  and we have  two  possibilities. If $R$ is  a real $2d$ IR, then TRS  applies  $(e_1,e_2)$ onto itself, with $e_i^*=e_i$. In this case, $\Theta=\mathbf{1}$ and Eq.~(\ref{invtime})  requires $H(-\vec k)=H(\vec k)^*$,  which is not satisfied by the Weyl hamiltonian. Thus $2d$ real IRs are excluded from Table~\ref{t1}. For complex and pseudoreal $2d$ IRs and for the real $4d$ $R_3$ of $\mathbf{212}$ and $\mathbf{213}$, $(e_1^*,e_2^*)$ are linearly independent of  $(e_1, e_2)$.  We can take as $4d$  basis either $(e_1, e_2,  e_1^*,  e_2^*)$ or $(e_1, e_2, i e_2^*, -i e_1^*)$, and everything proceeds as in the previous case. The main difference is that now 
 the off-diagonal blocks need not vanish by translation symmetry. However, a detailed case by case analysis using space group and TRS invariance shows that the off-diagonal terms are at least of  $O(k^2)$. This completes the proof of Eq.~(\ref{DW}).


\end{document}